\documentclass[aps,10pt,prd,notitlepage,showpacs,nofootinbib,superscriptaddress]{revtex4-2}
\usepackage{graphicx}
\usepackage[utf8]{inputenc} 
\usepackage{amsmath}
\usepackage{amssymb}
\usepackage{siunitx}
\usepackage[section]{placeins}
\usepackage{float}
\usepackage{comment}
\usepackage{slashed}
\usepackage[normalem]{ulem}
\usepackage{braket}
\usepackage{comment}

\usepackage{caption}
\usepackage{subcaption}
\usepackage[usenames,dvipsnames]{color}
\usepackage[colorinlistoftodos]{todonotes}
\usepackage[colorlinks=true,citecolor=blue,urlcolor=magenta, pdfborder={0 0 0}]{hyperref}

\usepackage{soul}

\usepackage[commandnameprefix=always]{changes}

\usepackage{slashed}

\usepackage{changes}
\begin{document}
\title{Dark Temperature Hierarchies and Gravitational Waves from the Electroweak Phase Transition}

\author{Arnab Chaudhuri}
\email{arnab.chaudhuri@vit.ac.in}
\affiliation{Department of Physics, School of Advanced Sciences,
Vellore Institute of Technology, Vellore, Tamil Nadu 632014, India.}

\begin{abstract}
We investigate the impact of a semi-decoupled dark sector with a temperature hierarchy relative to the Standard Model plasma on the electroweak phase transition and its associated gravitational wave signal. Working within a minimal Higgs-portal extension, we allow the dark sector to possess a higher temperature at the electroweak epoch while remaining consistent with cosmological bounds on additional relativistic degrees of freedom. The temperature hierarchy modifies the thermal structure of the effective potential and alters nucleation dynamics without requiring large portal couplings or extreme supercooling. Within the cosmologically allowed window, we find a monotonic enhancement of the gravitational wave amplitude by more than an order of magnitude compared to the standard thermal case, accompanied by a shift of the peak frequency within the millihertz regime. The resulting stochastic background moves substantially closer to the projected sensitivity of future space-based interferometers. Our results demonstrate that hidden-sector temperature hierarchies can leave observable imprints on electroweak-scale phase transitions even in minimal and perturbative frameworks.
\end{abstract}

\maketitle

%--------------------------------------------
\section{Introduction}
\label{sec:intro}

Phase transitions in the early Universe provide a powerful probe of physics beyond the Standard Model. In particular, a first-order electroweak phase transition (EWPT) would generate a stochastic gravitational wave background potentially observable by space-based interferometers operating in the millihertz frequency range \cite{Kosowsky:1992vn,Kamionkowski:1993fg,Hogan:1986qda,Caprini:2015zlo,Caprini:2019egz}. Because the Standard Model predicts a smooth crossover for the observed Higgs mass, additional degrees of freedom are required to realize a first-order transition \cite{Quiros:1999jp,Trodden:1998ym,Morrissey:2012db,Mazumdar:2018dfl}. Among the simplest and most extensively studied possibilities is the addition of a gauge-singlet scalar coupled to the Higgs through a portal interaction \cite{Espinosa:1993bs,Noble:2007kk,Profumo:2007wc,Espinosa:2008kw,Barger:2008jx,Curtin:2014jma,Chen:2014ask,Chaudhuri:2021agl,Chaudhuri:2021rwt,Chaudhuri:2021ppr,Chaudhuri:2021ibc}.

Singlet extensions have been widely explored in the context of electroweak baryogenesis and gravitational wave production \cite{Kozaczuk:2015owa,Huang:2017jws,Beniwal:2018hyi,Basler:2018cwe,Alanne:2019bsm,Croon:2020cgk,Chaudhuri:2025ybh,Chaudhuri:2022sis}. In conventional analyses, the hidden sector is assumed to share a common temperature with the Standard Model plasma during the electroweak epoch. Under this assumption, the thermal corrections entering the finite-temperature effective potential are entirely determined by the visible-sector bath, and the resulting gravitational wave signal depends sensitively on the detailed scalar spectrum and portal coupling.

However, the assumption of a common temperature is not generic. If a hidden sector decouples from the visible sector at sufficiently high temperatures, entropy conservation in each sector implies that the two temperatures evolve independently \cite{Kolb:1990vq,Hall:2009bx,Feng:2008mu,Chu:2011be,Blennow:2012de}. In such scenarios the dark-sector temperature at the electroweak epoch is not fixed a priori, but becomes an additional cosmological parameter constrained primarily by bounds on extra relativistic energy density. These bounds are commonly expressed in terms of $\Delta N_{\rm eff}$ and are tightly constrained by cosmic microwave background and big bang nucleosynthesis data \cite{Planck:2018jri,Planck:2018vyg,Cyburt:2015mya,Fields:2019pfx}. Hidden sectors with temperature hierarchies have been studied in a variety of cosmological contexts, including dark radiation, freeze-in dark matter, and non-standard thermal histories \cite{Berlin:2016gtr,Cui:2017ufi,Dror:2018pdh,Co:2015pka}.  

Despite the extensive literature on singlet-driven electroweak phase transitions and gravitational waves, the effect of a semi-decoupled temperature hierarchy on electroweak-scale gravitational wave phenomenology has not been systematically quantified. A hotter dark bath modifies the thermal corrections to the effective potential, altering both the quadratic and bosonic cubic temperature-dependent terms. Because these coefficients enter differently in the determination of the critical temperature, the nucleation action, and the inverse duration parameter, even moderate temperature hierarchies can significantly affect the latent heat parameter $\alpha$, the inverse duration $\beta/H$, and the nucleation temperature $T_n$. These quantities directly control the amplitude and peak frequency of the gravitational wave spectrum \cite{Huber:2008hg,Hindmarsh:2015qta,Hindmarsh:2017gnf,Weir:2017wfa,Chaudhuri:2022sis,Chaudhuri:2024vrd,Srivastava:2025oer,Chaudhuri:2025ybh,Chaudhuri:2025cjp,Chaudhuri:2025ylu,Witten:1984rs,Hogan:1986qda,Kosowsky:1991ua, Kosowsky:1992rz, Kamionkowski:1993fg, Caprini:2015zlo, Caprini:2019egz, Mazumdar:2018dfl, Caprini:2009fx, Hindmarsh:2017gnf,Athron:2023xlk,Ghosh:2022fzp,Bittar:2025lcr, Grojean:2006bp, Espinosa:2010hh, Patel:2011th, Kakizaki:2015wua, Huang:2017rzf, Athron:2019teq, Hashino:2018wee, Chala:2016ykx, Croon:2020cgk, Ramsey-Musolf:2019lsf, Fairbairn:2019xog, Ellis:2019oqb}.

Recent years have seen substantial progress in understanding gravitational wave production from cosmological phase transitions, including improved modeling of sound-wave dynamics and numerical simulations of bubble expansion \cite{Cutting:2018tjt,Caprini:2019egz,Croon:2020cgk,Das:2026zuo,Srivastava:2025oer}. Planned space-based interferometers such as LISA, BBO, and DECIGO provide strong motivation to revisit electroweak-scale transitions with refined theoretical tools \cite{LISA:2017pwj,LISA:2022yao,Kawamura:2011zz,Corbin:2005ny,Yagi:2011wg}. In this context, it is timely to reassess minimal scalar extensions under non-standard thermal histories that remain cosmologically viable.

In this work we investigate how a dark temperature hierarchy modifies the electroweak phase transition and its gravitational wave signal within a minimal Higgs-portal extension of the Standard Model. We allow the dark sector to possess a temperature $T_D = \xi T$ during the electroweak epoch and determine the cosmologically allowed range of $\xi$ consistent with bounds on $\Delta N_{\rm eff}$. We compute the finite-temperature effective potential including one-loop thermal corrections and daisy resummation, and evaluate bubble nucleation numerically using \texttt{CosmoTransitions} \cite{Wainwright:2011kj}. Focusing on perturbative portal couplings, we analyze how the phase transition parameters vary across the allowed hierarchy window and compute the resulting gravitational wave spectra and their signal-to-noise ratio for LISA.

We find that increasing the dark temperature ratio within cosmological bounds leads to a systematic enhancement of the gravitational wave amplitude by more than an order of magnitude, together with a shift of the peak frequency within the millihertz regime. The enhancement arises from modified nucleation dynamics induced by the hotter dark bath and occurs within a perturbative and cosmologically consistent framework. Our results demonstrate that hidden-sector temperature hierarchies can leave observable imprints on electroweak-scale phase transitions even in minimal scalar portal models.

The paper is organized as follows. In Sec.~\ref{sec:model} we introduce the Higgs-portal framework and discuss the cosmological consistency conditions allowing a temperature hierarchy between the dark and visible sectors. In Sec.~\ref{sec:pt} we analyze the electroweak phase transition dynamics and present the numerical results for the nucleation parameters. Sec.~\ref{sec:gw} is devoted to the resulting gravitational wave spectra and their comparison with projected detector sensitivities. We conclude in Sec.~\ref{sec:discussion}.

\section{Model and cosmological consistency}
\label{sec:model}

We consider the Standard Model (SM) extended by a real scalar singlet $S$ interacting through a Higgs portal coupling. Such extensions constitute one of the minimal frameworks for modifying electroweak phase transition dynamics while preserving the SM gauge structure \cite{Patt:2006fw,Espinosa:2011ax,Curtin:2014jma}.

We impose a discrete $\mathbb{Z}_2$ symmetry under which
\begin{equation}
S \to -S ,
\end{equation}
while all Standard Model fields remain even. This symmetry forbids linear and cubic singlet operators and ensures that $S=0$ is an extremum of the tree-level potential.

The scalar-sector Lagrangian is
\begin{equation}
\mathcal{L}
=
\mathcal{L}_{\rm SM}
+
\frac{1}{2}(\partial_\mu S)^2
-
\frac{1}{2} m_{S0}^2 S^2
-
\frac{\lambda_S}{4} S^4
-
\frac{\kappa}{2}\, H^\dagger H\, S^2 ,
\label{eq:lagrangian}
\end{equation}
where $m_{S0}$ is the singlet mass parameter, $\lambda_S$ the singlet quartic coupling, and $\kappa$ the portal interaction strength.

Electroweak symmetry breaking proceeds along the Higgs direction provided
\begin{equation}
m_{S0}^2 + \frac{\kappa}{2} v^2 > 0 ,
\end{equation}
so that the singlet does not acquire a vacuum expectation value in the broken phase.

In unitary gauge,
\[
H = \frac{1}{\sqrt{2}}
\begin{pmatrix}
0\\ h
\end{pmatrix},
\]
where $h$ denotes the neutral Higgs field. The tree-level Higgs potential is
\begin{equation}
V_0(h)
=
-\frac{1}{2}\mu^2 h^2
+
\frac{1}{4}\lambda h^4 ,
\end{equation}
with $\lambda \simeq 0.13$ fixed by $m_h=125$ GeV and $v=246$ GeV. Along the Higgs background, the singlet field-dependent mass is
\begin{equation}
m_S^2(h)
=
m_{S0}^2
+
\frac{\kappa}{2} h^2 .
\label{eq:mSh}
\end{equation}

The electroweak phase transition is analyzed using the full one-loop finite-temperature effective potential including Coleman--Weinberg corrections, thermal contributions, daisy resummation, and counterterms. Explicit expressions and renormalization conditions are summarized in Appendix~\ref{app:Veff}.

For analytic intuition, the high-temperature expansion admits the approximate cubic form
\begin{equation}
V_{\rm eff}(h,T)
=
D_\xi (T^2 - T_0^2) h^2
-
E_\xi T h^3
+
\frac{\lambda_T}{4} h^4 .
\label{eq:Veff}
\end{equation}
The cubic term arises from the bosonic zero Matsubara modes in the high-temperature expansion of the thermal integrals and generates the barrier between phases. It originates from gauge bosons and the singlet scalar and does not violate the $\mathbb{Z}_2$ symmetry, since it is induced radiatively at finite temperature in the Higgs direction.

We allow the singlet sector to possess a temperature $T_D$ different from the SM temperature $T$, defining
\[
\xi \equiv \frac{T_D}{T}.
\]
In the high-temperature limit,
\begin{align}
D_\xi &= D_{\rm SM} + \frac{\kappa}{24}\xi^2 , \\
E_\xi &= E_{\rm SM}
+
\frac{1}{12\pi}
\left(\frac{\kappa}{2}\right)^{3/2}\xi .
\end{align}
The $\xi^2$ scaling originates from quadratic thermal corrections proportional to $T_D^2$, while the linear $\xi$ dependence arises from the bosonic zero-mode contribution to the cubic term \cite{Quiros:1999jp}. Numerically, $D_{\rm SM}\simeq 0.4$ and $E_{\rm SM}\simeq 0.01$.

The singlet thermal mass entering ring resummation is
\begin{equation}
\Pi_S(T_D)
=
\frac{\kappa}{12}\xi^2 T^2 ,
\end{equation}
which regulates infrared divergences from the zero mode \cite{Arnold:1992rz}. All numerical results below are obtained from the full one-loop resummed potential.

A temperature hierarchy requires that the singlet sector is not in thermal equilibrium with the SM plasma at the electroweak epoch. The dominant portal-induced process is $HH \leftrightarrow SS$. At $T \gg m_i$,
\[
\sigma \sim \frac{\kappa^2}{T^2},
\qquad
\Gamma \sim \kappa^2 T .
\]
Comparing with the Hubble rate during radiation domination,
\[
H(T)
=
\sqrt{\frac{\pi^2 g_*}{90}}
\frac{T^2}{M_{\rm Pl}},
\]
decoupling occurs when $\Gamma \lesssim H$, giving parametrically
\begin{equation} \label{eq:Tdec}
T_{\rm dec}
\sim
\frac{\kappa^2}{\sqrt{g_*}} M_{\rm Pl}.
\end{equation}
For $\kappa \sim \mathcal{O}(1)$, this scale is far above the electroweak temperature, ensuring thermal decoupling during the phase transition. In deriving Eq.~\eqref{eq:Tdec} we have assumed relativistic scattering in the symmetric phase, valid for temperatures well above the electroweak scale. For portal couplings in the perturbative regime, $\kappa \lesssim \mathcal{O}(1)$, the decoupling temperature obtained from $\Gamma \sim H$ lies parametrically far above the electroweak scale, ensuring that the singlet sector is thermally separated prior to the phase transition. Once decoupled, entropy conservation in each sector independently preserves the temperature ratio $\xi$, up to the standard dilution from changes in the effective number of relativistic degrees of freedom.

At temperatures near and below the electroweak scale, the Higgs field acquires a vacuum expectation value and both Higgs and singlet states become massive. The number densities entering the portal-induced interaction rate $\Gamma \sim n \langle \sigma v \rangle$ are then Boltzmann suppressed, causing $\Gamma$ to decrease more rapidly than the Hubble rate. Consequently, thermal re-equilibration does not occur during or after the electroweak phase transition. The temperature hierarchy therefore remains stable throughout the transition in the perturbative parameter region considered in this work.

Entropy conservation implies
\begin{equation}
\xi_{\rm BBN}
=
\xi_{\rm EW}
\left(
\frac{g_{*,s}(T_{\rm BBN})}{g_{*,s}(T_{\rm dec})}
\right)^{1/3}.
\end{equation}
Using $g_{*,s}(T_{\rm dec})\simeq106.75$ and $g_{*,s}(T_{\rm BBN})\simeq10.75$, the dilution factor is approximately $0.46$.

A relativistic real scalar contributes
\begin{equation}
\Delta N_{\rm eff}
=
\frac{4}{7}
\left(\frac{11}{4}\right)^{4/3}
\xi_{\rm BBN}^4 ,
\end{equation}
which must satisfy $\Delta N_{\rm eff}\lesssim0.3$ \cite{Planck:2018jri,Cyburt:2015mya}. This implies
\begin{equation}
\xi_{\rm EW} \lesssim 1.8 .
\end{equation}

We therefore consider
\[
1.0 \le \xi \lesssim 1.8 ,
\]
with perturbative portal couplings.

\subsection{Benchmark parameter choice and theoretical consistency}
\label{subsec:benchmark}

The Standard Model parameters are fixed by
\[
m_h = 125~{\rm GeV},
\qquad
v = 246~{\rm GeV},
\]
which determine $\lambda \simeq 0.13$ at the renormalization scale $\mu_R = v$.

We fix
\[
\kappa = 1,
\qquad
\lambda_S = 0.5,
\]
both within the perturbative regime.

We define the physical singlet mass in the broken phase as
\begin{equation}
m_S^2 = m_{S0}^2 + \frac{\kappa}{2} v^2 .
\end{equation}
Throughout this work we adopt
\[
m_S = 250~{\rm GeV},
\]
which avoids Higgs decay constraints and is consistent with current collider limits for a $\mathbb{Z}_2$-odd singlet. This benchmark value is chosen for several phenomenological and dynamical reasons. 
First, $m_S > m_h/2$ ensures that the decay $h \to SS$ is kinematically forbidden, 
avoiding constraints from the Higgs invisible width for portal couplings of order unity. 
Second, for an exact $\mathbb{Z}_2$ symmetry the singlet does not mix with the Higgs, 
and current collider searches place only weak direct limits on a $\mathbb{Z}_2$-odd scalar 
with mass in the few-hundred-GeV range and perturbative portal coupling. 
Third, taking $m_S$ to be of the same order as the electroweak scale 
ensures that the singlet contributes appreciably to thermal corrections in the effective potential 
without decoupling from the plasma dynamics or inducing extreme supercooling. 

We have verified that moderate variations of $m_S$ within the range 
$200\text{–}400~{\rm GeV}$ do not qualitatively modify the hierarchy-induced enhancement mechanism discussed below. 
The choice $m_S = 250~{\rm GeV}$ should therefore be understood as a representative 
benchmark within a broader perturbative parameter region rather than a finely tuned point.

For $\kappa=1$, this gives
\[
m_{S0}^2
=
m_S^2 - \frac{1}{2}v^2
\simeq (180~{\rm GeV})^2 .
\]

Vacuum stability requires
\[
\lambda > 0,
\qquad
\lambda_S > 0,
\qquad
\kappa > -2\sqrt{\lambda\lambda_S},
\]
which are satisfied. Perturbativity is maintained since $\kappa,\lambda_S \ll 4\pi$.

With the scalar sector specified, the only remaining continuous parameter controlling the thermal dynamics is the temperature ratio $\xi=T_D/T$. All phase transition and gravitational wave results presented below follow from this fully specified and internally consistent benchmark realization.

\section{Electroweak phase transition dynamics}
\label{sec:pt}

We now analyze the electroweak phase transition in the presence of a dark temperature hierarchy. The quantities characterizing the transition are the critical temperature $T_c$, the nucleation temperature $T_n$, the latent heat parameter $\alpha$, and the inverse duration parameter $\beta/H$ \cite{Quiros:1999jp,Caprini:2015zlo}.

The critical temperature is defined by the degeneracy condition between the symmetric minimum at $h=0$ and the broken minimum at $h\neq 0$. Using the cubic parametrization of Eq.~\eqref{eq:Veff}, this condition yields
\begin{equation}
T_c^2
=
\frac{\mu^2}
{2D_\xi - 2E_\xi^2/\lambda_T} ,
\label{eq:Tc}
\end{equation}
valid within the high-temperature approximation. This expression illustrates how the thermal coefficients $D_\xi$ and $E_\xi$ control the onset of symmetry breaking. Since $D_\xi$ grows as $\xi^2$ while $E_\xi$ grows linearly in $\xi$, increasing the dark temperature ratio modifies the balance between the quadratic and cubic terms, leading to a systematic shift in $T_c$. For the benchmark choice $\kappa=1$, a numerical determination based on the full one-loop resummed potential yields $T_c$ decreases from $96.5$ GeV at $\xi=1$ to $85.5$ GeV at $\xi=1.8$.

The phase transition completes when thermal tunnelling becomes efficient. The nucleation temperature $T_n$ is defined by the condition
\begin{equation}
\frac{S_3(T_n)}{T_n} \simeq 140 ,
\label{eq:nucleation_condition}
\end{equation}
which corresponds to approximately one bubble nucleated per Hubble volume in a radiation-dominated Universe \cite{Quiros:1999jp}. Here $S_3$ denotes the three-dimensional Euclidean action of the bounce solution computed along the Higgs direction. We compute $S_3(T)$ numerically using \texttt{CosmoTransitions}, solving the overshoot/undershoot boundary value problem for the full one-loop finite-temperature effective potential including ring resummation. In particular, the tunnelling calculation does not rely on the cubic approximation of Eq.~\eqref{eq:Veff}.

The strength and duration of the transition are characterized by the parameters
\begin{equation}
\alpha
=
\frac{1}{\rho_{\rm rad}}
\left[
\Delta V
-
T \frac{d\Delta V}{dT}
\right]_{T=T_n},
\qquad
\frac{\beta}{H}
=
T \left.
\frac{d}{dT}
\left(\frac{S_3}{T}\right)
\right|_{T=T_n},
\label{eq:alpha_beta}
\end{equation}
where $\Delta V \equiv V_{\rm eff}(h_{\rm false},T) - V_{\rm eff}(h_{\rm true},T)$ is the difference in free energy density between the false and true vacua at temperature $T$, and $\rho_{\rm rad}=\frac{\pi^2}{30}g_*T_n^4$ is the radiation energy density with $g_*=106.75$. The parameter $\alpha$ measures the fraction of vacuum energy released relative to the radiation background, while $\beta/H$ quantifies the inverse duration of the transition in units of the Hubble rate \cite{Caprini:2015zlo,Caprini:2019egz}.

We fix $\kappa=1$ and vary $\xi$ within the cosmologically allowed window $1 \le \xi \le 1.8$. The numerical results obtained from the full one-loop potential are summarized in Table~\ref{tab:benchmarks}.

\begin{table}[t]
\centering
\begin{tabular}{c c c c c}
\hline
$\xi$ & $T_c$ (GeV) & $T_n$ (GeV) & $\alpha$ & $\beta/H$ \\
\hline
1.0 & 96.55 & 89.23 & 0.046 & 308 \\
1.2 & 93.20 & 86.19 & 0.0658 & 256 \\
1.4 & 90.04 & 83.01 & 0.079 & 202 \\
1.6 & 88.01 & 78.54 & 0.085 & 176 \\
1.8 & 85.48 & 75.01 & 0.109 & 105 \\
\hline
\end{tabular}
\caption{Phase transition parameters for $\kappa=1$ and varying dark temperature ratio $\xi=T_D/T$.}
\label{tab:benchmarks}
\end{table}

Several systematic trends emerge from the numerical analysis. The nucleation temperature decreases monotonically as $\xi$ increases, reflecting the modified thermal contributions to the effective potential. Simultaneously, the latent heat parameter $\alpha$ increases across the allowed hierarchy window, while the inverse duration parameter $\beta/H$ decreases substantially. Between $\xi=1$ and $\xi=1.8$, $\alpha$ increases by more than a factor of two and $\beta/H$ decreases by nearly a factor of three.

The ratio $T_n/T_c$ correspondingly decreases with increasing $\xi$, indicating progressively stronger supercooling within the cosmologically allowed window, although the transition remains within a perturbative regime.

These correlated shifts in $\alpha$ and $\beta/H$ play a central role in determining the amplitude and peak structure of the stochastic gravitational wave background generated by the transition, which we analyze in the following section.

\section{Gravitational wave signal}
\label{sec:gw}

A first-order electroweak phase transition generates a stochastic gravitational wave background through three distinct mechanisms: scalar field gradients from bubble wall collisions, long-lasting acoustic waves in the plasma, and magnetohydrodynamic turbulence \cite{Kosowsky:1992vn,Kosowsky:1992rz,Caprini:2015zlo,Caprini:2019egz}.  The total energy density may be schematically decomposed as
\begin{equation}
\Omega_{\rm GW}(f)
=
\Omega_{\rm col}(f)
+
\Omega_{\rm sw}(f)
+
\Omega_{\rm turb}(f),
\end{equation}
where $\Omega_{\rm col}$ denotes the envelope (collision) contribution, $\Omega_{\rm sw}$ the acoustic component, and $\Omega_{\rm turb}$ the turbulent contribution.  

For transitions in which bubble walls do not run away, numerical simulations indicate that the sound-wave contribution dominates the total signal, while the collision term is suppressed and turbulence provides a subleading correction \cite{Hindmarsh:2015qta,Hindmarsh:2017gnf,Caprini:2019egz}.  In the parameter region considered here, characterized by moderate $\alpha$ and $\beta/H \gtrsim 100$, runaway behavior is not expected.  We therefore focus on the acoustic contribution as the leading and most robust source of gravitational radiation and neglect $\Omega_{\rm col}$ and $\Omega_{\rm turb}$ in the quantitative predictions below.

\begin{figure}[t]
\centering
\includegraphics[width=0.9\textwidth]{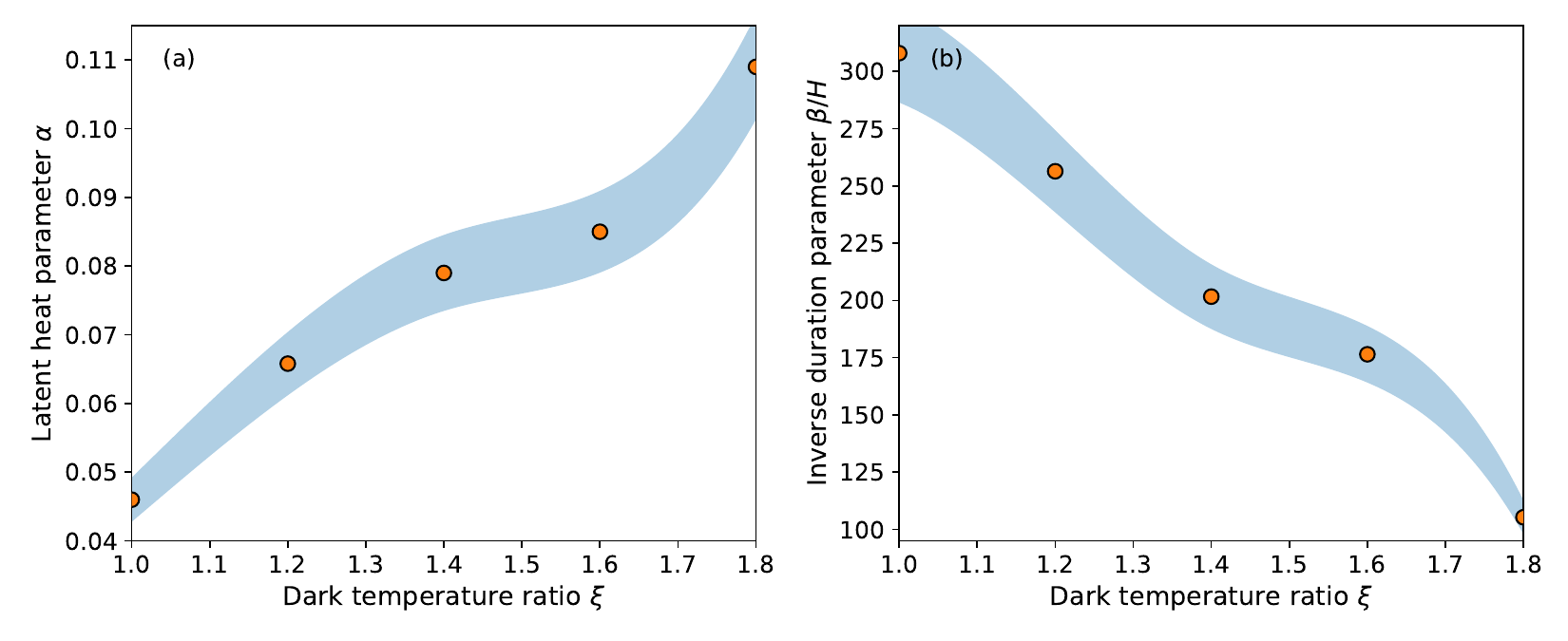}
\caption{
Dependence of (a) the latent heat parameter $\alpha$ and (b) the inverse duration parameter $\beta/H$ on the dark temperature ratio $\xi=T_D/T$ for $\kappa=1$.  Points denote numerical evaluations using the full one-loop effective potential with ring resummation.
}
\label{fig:alpha_beta}
\end{figure}

Before presenting the predicted spectra, it is useful to summarize how the temperature hierarchy modifies the macroscopic transition parameters entering the signal calculation.  Figure~\ref{fig:alpha_beta} displays the dependence of the latent heat parameter $\alpha$ and the inverse duration parameter $\beta/H$ on the dark temperature ratio $\xi$ for $\kappa=1$.  As $\xi$ increases within the cosmologically allowed window, $\alpha$ increases monotonically while $\beta/H$ decreases substantially.  Since the gravitational wave amplitude scales with $H_*/\beta$ and with the released vacuum energy fraction, these correlated shifts directly control both the normalization and the peak frequency of the signal.

The present-day gravitational wave energy density from long-lasting acoustic waves can be expressed as \cite{Caprini:2015zlo,Caprini:2019egz}
\begin{equation}
\Omega_{\rm sw}(f) h^2
=
2.65 \times 10^{-6}
\left(\frac{H_*}{\beta}\right)
\left(\frac{\kappa_v \alpha}{1+\alpha}\right)^2
\left(\frac{100}{g_*}\right)^{1/3}
v_w
\, S_{\rm sw}(f),
\label{eq:OmegaGW_full}
\end{equation}
where $H_*$ is the Hubble rate at $T_n$, $\alpha$ is the ratio of released vacuum energy to radiation energy density, $\beta/H_*$ characterizes the inverse duration of the transition, $g_*$ is the number of relativistic degrees of freedom, $v_w$ is the bubble wall velocity, and $\kappa_v$ denotes the fraction of vacuum energy converted into bulk fluid motion.

We adopt the standard hydrodynamic fit appropriate for non-runaway walls,
\begin{equation}
\kappa_v \simeq
\frac{\alpha}{0.73 + 0.083\sqrt{\alpha} + \alpha}.
\end{equation}

Throughout this work we assume a representative subsonic wall velocity
\[
v_w = 0.6,
\]
consistent with deflagration solutions in moderately strong transitions and commonly adopted in phenomenological studies.

The spectral shape function is given by
\begin{equation}
S_{\rm sw}(f)
=
\left(\frac{f}{f_{\rm peak}}\right)^3
\left[
\frac{7}{4 + 3 (f/f_{\rm peak})^2}
\right]^{7/2},
\label{eq:SpectralShape}
\end{equation}
which reproduces the characteristic $f^3$ infrared rise and power-law decay at higher frequencies.

The peak frequency today is approximately
\begin{equation}
f_{\rm peak}
\simeq
1.9 \times 10^{-5} \,{\rm Hz}\,
\frac{1}{v_w}
\left(\frac{\beta}{H_*}\right)
\left(\frac{T_n}{100\,{\rm GeV}}\right)
\left(\frac{g_*}{100}\right)^{1/6}.
\label{eq:fpeak_full}
\end{equation}

Parametrically, the peak amplitude scales as
\[
\Omega_{\rm peak}
\sim
\left(\frac{H_*}{\beta}\right)
\left(\frac{\alpha}{1+\alpha}\right)^2,
\]
up to order-one factors associated with $v_w$ and the spectral shape.  Increasing $\alpha$ while decreasing $\beta/H_*$ therefore coherently enhances the signal.

\begin{figure}[t]
\centering
\includegraphics[width=0.85\textwidth]{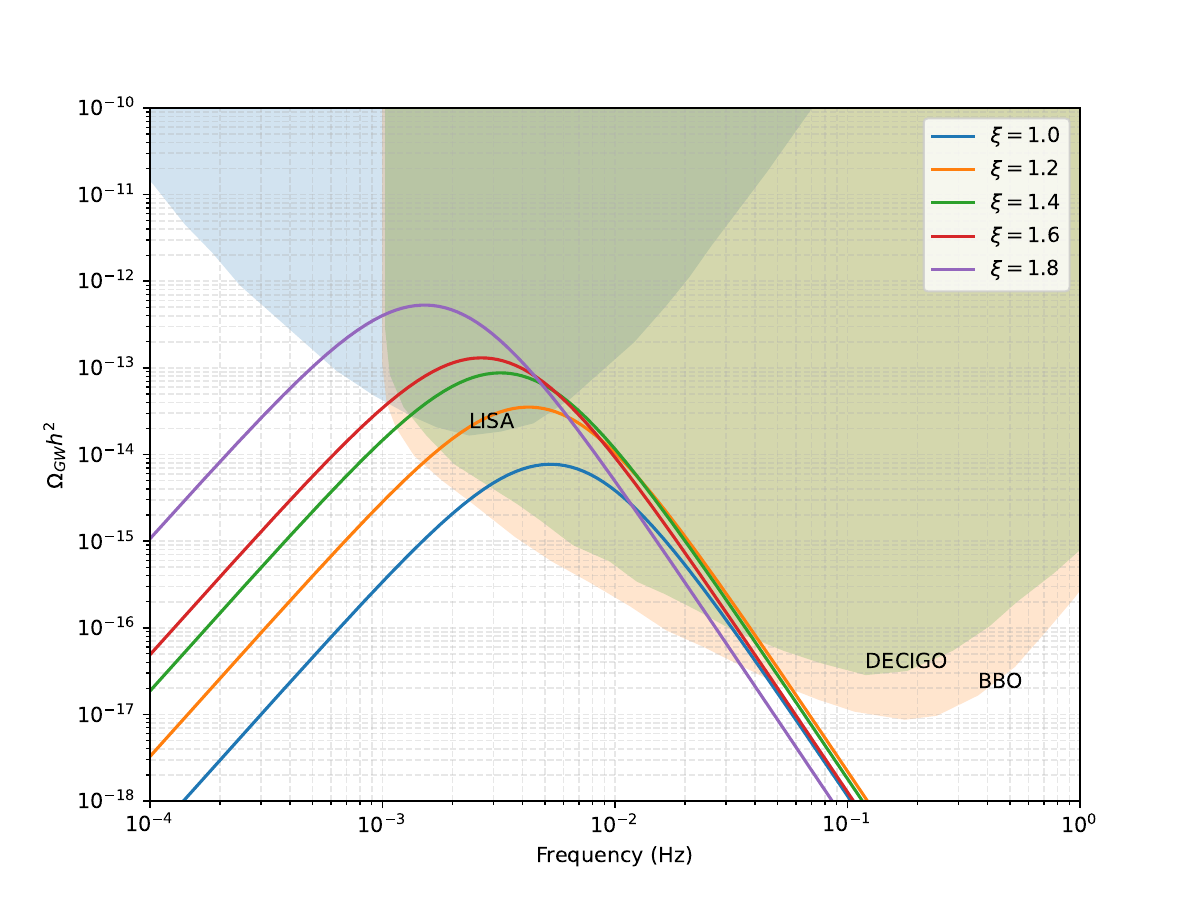}
\caption{
Gravitational wave spectra for $\kappa=1$ and varying dark temperature ratio $\xi=T_D/T$.  Increasing $\xi$ enhances the amplitude and shifts the peak toward lower frequencies.
}
\label{fig:GW}
\end{figure}

Figure~\ref{fig:GW} displays the resulting spectra for the benchmark points listed in Table~\ref{tab:benchmarks}.  Across the allowed hierarchy window, $T_n$ decreases from $89$ GeV to $75$ GeV, $\alpha$ increases from $0.046$ to $0.109$, and $\beta/H$ decreases from $308$ to $105$.  These combined effects yield approximately a factor of $\sim 10$ enhancement of the peak amplitude between $\xi=1$ and $\xi=1.8$.  The peak frequency remains in the millihertz regime, $f_{\rm peak}\sim 10^{-3}$--$10^{-2}$ Hz, characteristic of electroweak-scale transitions.

The enhancement mechanism arises from modified thermal coefficients induced by a hotter dark sector, which alter the nucleation dynamics within a controlled perturbative regime.

Finally, Eq.~\eqref{eq:OmegaGW_full} assumes that acoustic modes persist for approximately a Hubble time.  Recent numerical simulations indicate that finite sound-wave lifetimes may introduce an $\mathcal{O}(1)$ suppression in the absolute normalization \cite{Hindmarsh:2017gnf,Caprini:2019egz}.  Such uncertainties affect the overall amplitude uniformly and do not modify the relative enhancement induced by the temperature hierarchy.  The monotonic amplification with increasing $\xi$ therefore remains a robust qualitative prediction of the framework.

\subsection{Signal-to-noise ratio and detectability}
\label{subsec:snr}

The gravitational wave spectra shown in Fig.~\ref{fig:GW} illustrate the hierarchy-induced enhancement.  Experimental observability is quantified through the signal-to-noise ratio (SNR).  For a stochastic background observed over time $T_{\rm obs}$, the SNR is given by \cite{Thrane:2013oya,Caprini:2019egz}
\begin{equation}
{\rm SNR}^2
=
T_{\rm obs}
\int_{f_{\rm min}}^{f_{\rm max}}
df \,
\left[
\frac{\Omega_{\rm GW}(f)}
{\Omega_{\rm sens}(f)}
\right]^2 ,
\label{eq:SNR}
\end{equation}
where $\Omega_{\rm sens}(f)$ denotes the detector sensitivity curve expressed in energy-density units.

For LISA we adopt $T_{\rm obs}=4~{\rm yr}$ and use the publicly available sensitivity including instrumental and confusion noise.  Electroweak-scale transitions produce millihertz peaks (cf.~Eq.~\eqref{eq:fpeak_full}), placing the signal squarely within the LISA band.

\begin{figure}[t]
\centering
\includegraphics[width=0.8\textwidth]{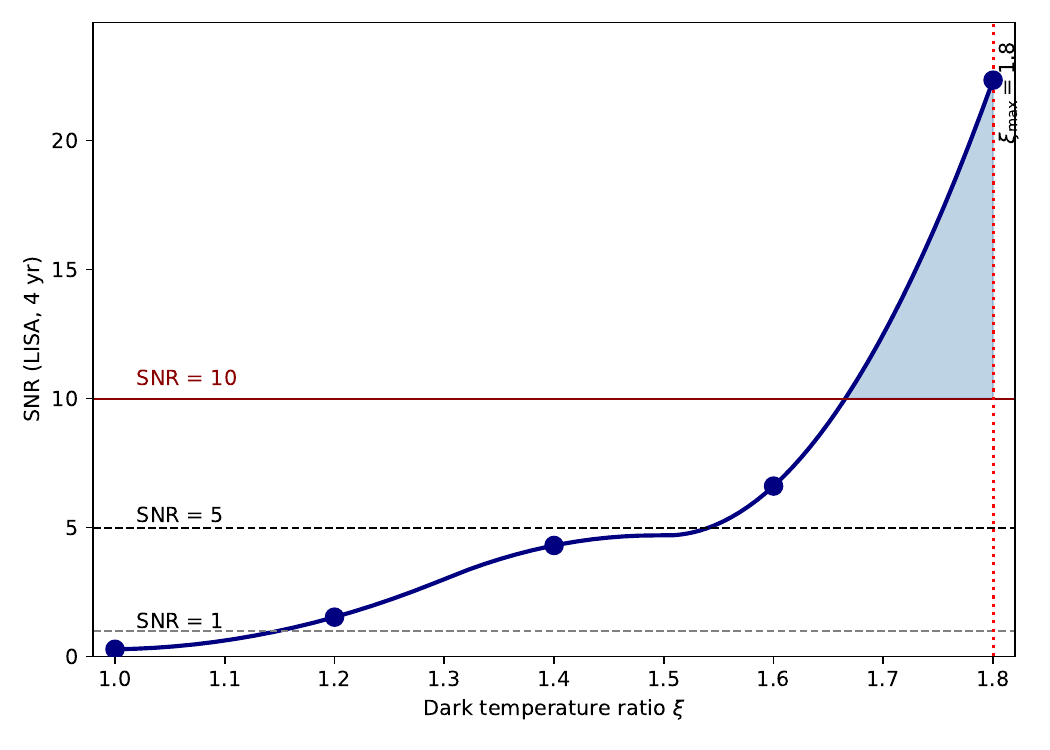}
\caption{
Signal-to-noise ratio for LISA (four-year mission) as a function of the dark temperature ratio $\xi$ for $\kappa=1$.  
The shaded region indicates ${\rm SNR} \ge 10$.  
The vertical dashed line marks the cosmological bound $\xi_{\rm max}=1.8$.}
\label{fig:SNR}
\end{figure}

The resulting SNR as a function of $\xi$ for $\kappa=1$ is shown in Fig.~\ref{fig:SNR}.  The shaded region corresponds to ${\rm SNR}\ge 10$, a conservative criterion for robust detection.

The SNR increases steeply with $\xi$, reflecting the correlated evolution of $\alpha$ and $\beta/H$.  The maximal benchmark yields ${\rm SNR}\simeq 22$, demonstrating that within cosmologically allowed bounds a moderate temperature hierarchy can elevate an otherwise marginal electroweak signal into a robustly detectable regime.

Uncertainties in the sound-wave lifetime may introduce order-one normalization shifts \cite{Hindmarsh:2017gnf}, but do not alter the monotonic growth with $\xi$.  The qualitative conclusion remains stable.

\section{Discussion and conclusions}
\label{sec:discussion}

We have investigated the impact of a semi-decoupled dark sector with temperature hierarchy $T_D/T \equiv \xi$ on the electroweak phase transition and its associated gravitational wave signal.  The framework is deliberately minimal: the Standard Model is extended by a single real scalar singlet coupled through a Higgs portal interaction with perturbative strength $\kappa=1$.  The dark sector is allowed to possess a higher temperature than the visible plasma at the electroweak epoch while remaining consistent with cosmological bounds on $\Delta N_{\rm eff}$.

The central result is that a temperature hierarchy modifies the thermal coefficients of the finite-temperature effective potential in a controlled manner, thereby altering nucleation dynamics without requiring large portal couplings or extreme supercooling.  As the ratio $\xi$ increases within the cosmologically allowed window $1 \le \xi \lesssim 1.8$, we find a monotonic decrease in $\beta/H$, a systematic increase in the latent heat parameter $\alpha$, and a corresponding reduction in the nucleation temperature.  These correlated effects enhance the peak gravitational wave amplitude by $\mathcal{O}(10)$ and shift the signal toward lower frequencies at a few $\times 10^{-3}$ Hz.

To quantify experimental prospects, we compute the signal-to-noise ratio for LISA assuming a four-year mission.  While the thermally equilibrated case $\xi=1$ yields ${\rm SNR}<1$, increasing the temperature hierarchy rapidly enhances detectability.  The SNR exceeds ${\rm SNR}=5$ for $\xi \gtrsim 1.6$ and reaches ${\rm SNR}\simeq 22$ near the cosmological upper bound $\xi\simeq 1.8$, corresponding to robust detection under conservative criteria.  Within cosmologically allowed limits, a moderate temperature hierarchy is therefore sufficient to move an otherwise marginal electroweak signal into the observable regime of LISA.

The enhancement mechanism arises from modified thermal contributions to the quadratic and cubic terms of the effective potential induced by the hotter dark bath.  The analysis remains within a perturbative portal regime and satisfies conservative bounds from $\Delta N_{\rm eff}$ after entropy dilution.  Importantly, the improvement in detectability does not rely on runaway bubble walls or extreme supercooling, but instead follows from controlled modifications of nucleation dynamics.

The absolute normalization of the sound-wave contribution carries theoretical uncertainties associated with the finite lifetime of acoustic modes in the plasma.  Numerical simulations indicate potential $\mathcal{O}(1)$ corrections to the amplitude \cite{Hindmarsh:2017gnf,Caprini:2019egz}.  Such uncertainties rescale the overall SNR but do not alter its monotonic growth with $\xi$, and therefore do not affect the qualitative conclusion regarding enhanced detectability.

More broadly, this study demonstrates that semi-decoupled dark sectors can materially influence electroweak-scale gravitational wave phenomenology even in minimal scalar portal extensions.  Cosmological temperature hierarchies—often neglected in phase transition analyses—can play a nontrivial role in determining nucleation dynamics and observable signatures.

Several directions merit further investigation.  A broader parameter scan including variation of the portal coupling and singlet self-interaction would clarify the maximal achievable enhancement within perturbative limits.  Embedding the scenario in ultraviolet-complete dark sector models could connect the temperature hierarchy to specific reheating or decoupling mechanisms.  Finally, improved simulations of acoustic dynamics would reduce the theoretical uncertainty in the predicted gravitational wave amplitude and sharpen forecasts for LISA.

In summary, a hotter dark bath consistent with cosmological bounds can enhance the electroweak gravitational wave signal by $\mathcal{O}(10)$ relative to the thermally equilibrated case.  For temperature hierarchies near the cosmological upper limit, the resulting signal achieves robust detectability at LISA, illustrating that hidden-sector thermal histories can leave observable imprints on cosmological phase transitions.

\appendix
\section{One-loop finite-temperature effective potential}
\label{app:Veff}

In this appendix we summarize the explicit form of the effective potential used in the numerical analysis.  All phase transition quantities are computed from the full one-loop finite-temperature effective potential with ring (daisy) resummation in Landau gauge within the $\overline{\rm MS}$ renormalization scheme.

The effective potential is written as
\begin{equation}
V_{\rm eff}(h,T)
=
V_0(h)
+
V_{\rm CW}(h)
+
V_T(h,T)
+
V_{\rm daisy}(h,T)
+
V_{\rm ct}(h),
\label{eq:Veff_full}
\end{equation}
where the individual contributions are described below.

%------------------------------------------------
\subsection{Tree-level scalar potential and field-dependent masses}
\label{subsec:treelevel}

The tree-level scalar potential in the background fields $(h,S)$ is
\begin{equation}
V_0(h,S)
=
-\frac{1}{2}\mu^2 h^2
+
\frac{1}{4}\lambda h^4
+
\frac{1}{2} m_{S0}^2 S^2
+
\frac{\lambda_S}{4} S^4
+
\frac{\kappa}{4} h^2 S^2 ,
\label{eq:Vtree_full}
\end{equation}
where $h$ denotes the neutral Higgs field in unitary gauge,
$H = (0,h/\sqrt{2})^T$.  

Along the Higgs direction $S=0$, the potential reduces to
\begin{equation}
V_0(h)
=
-\frac{1}{2}\mu^2 h^2
+
\frac{1}{4}\lambda h^4 .
\label{eq:Vtree_h}
\end{equation}

The field-dependent masses entering the one-loop effective potential are obtained by expanding around the background Higgs field \cite{Quiros:1999jp,Espinosa:2011ax}.  They are given by

\begin{align}
m_h^2(h) &= -\mu^2 + 3\lambda h^2 , \\
m_G^2(h) &= -\mu^2 + \lambda h^2 , \\
m_W^2(h) &= \frac{g^2}{4} h^2 , \\
m_Z^2(h) &= \frac{g^2+g'^2}{4} h^2 , \\
m_t^2(h) &= \frac{y_t^2}{2} h^2 , \\
m_S^2(h) &= m_{S0}^2 + \frac{\kappa}{2} h^2 .
\label{eq:field_masses}
\end{align}

Here $m_G$ denotes the Goldstone modes, $g$ and $g'$ are the electroweak gauge couplings, and $y_t$ is the top Yukawa coupling.  

The corresponding numbers of degrees of freedom contributing to the effective potential are

\begin{align}
n_h &= 1, \\
n_G &= 3, \\
n_W &= 6, \\
n_Z &= 3, \\
n_t &= -12, \\
n_S &= 1,
\end{align}

where the negative sign for fermions accounts for Fermi–Dirac statistics.  
For gauge bosons, these multiplicities correspond to transverse plus longitudinal polarizations at zero temperature.

These masses and multiplicities are used in the Coleman–Weinberg potential and in the finite-temperature corrections discussed below.

%------------------------------------------------
\subsection{Coleman--Weinberg potential}

The one-loop zero-temperature correction is given by the Coleman--Weinberg potential,
\begin{equation}
V_{\rm CW}(h)
=
\sum_i
\frac{n_i}{64\pi^2}
m_i^4(h)
\left[
\ln\!\left(\frac{m_i^2(h)}{\mu_R^2}\right)
-
c_i
\right],
\label{eq:CW}
\end{equation}
where $\mu_R$ is the renormalization scale.  In the $\overline{\rm MS}$ scheme,
\begin{equation}
c_i =
\begin{cases}
\frac{3}{2} & \text{(scalars and fermions)},\\
\frac{5}{6} & \text{(gauge bosons)}.
\end{cases}
\end{equation}
All couplings entering Eq.~\eqref{eq:CW} are evaluated at the renormalization scale $\mu_R = v = 246~{\rm GeV}$, corresponding to the physical electroweak vacuum expectation value.

%------------------------------------------------
\subsection{Finite-temperature corrections}

The finite-temperature contribution is
\begin{equation}
V_T(h,T)
=
\frac{T^4}{2\pi^2}
\sum_i
n_i
J_{\pm}\!\left(\frac{m_i^2(h)}{T^2}\right),
\label{eq:VT}
\end{equation}
where $J_+$ and $J_-$ correspond to bosons and fermions respectively,
\begin{equation}
J_{\pm}(y^2)
=
\int_0^\infty dx\, x^2
\ln\!\left[
1 \mp \exp\!\left(-\sqrt{x^2+y^2}\right)
\right].
\end{equation}

In the high-temperature limit, the bosonic integral contains a cubic term proportional to $m_i^3 T$, which is responsible for the barrier between phases through the contribution of bosonic zero Matsubara modes.

%------------------------------------------------
\subsection{Daisy (ring) resummation}

Infrared divergences from bosonic zero modes are regulated by ring resummation \cite{Arnold:1992rz,Parwani:1991gq}.  In practice, this amounts to replacing
\begin{equation}
m_i^2(h)
\rightarrow
m_i^2(h) + \Pi_i(T)
\end{equation}
in the bosonic thermal functions for the longitudinal degrees of freedom.

The leading thermal masses are

\begin{align}
\Pi_h(T) &= T^2\left(\frac{3g^2 + g'^2}{16} + \frac{y_t^2}{4} + \frac{\lambda}{2} + \frac{\kappa}{12}\xi^2 \right), \\
\Pi_S(T_D) &= \frac{\kappa}{12} T_D^2 = \frac{\kappa}{12}\xi^2 T^2 , \\
\Pi_W(T) &= \frac{11}{6} g^2 T^2 , \\
\Pi_Z(T) &= \frac{11}{6} (g^2 + g'^2) T^2 .
\end{align}

Only bosonic longitudinal modes receive thermal mass corrections; transverse gauge modes do not acquire Debye masses at leading order.

%------------------------------------------------
\subsection{Counterterm potential and renormalization}

To preserve the physical Higgs vacuum expectation value and mass at zero temperature, we include a counterterm potential.  
In the most general form consistent with the scalar sector,
\begin{equation}
V_{\rm ct}(h,S)
=
\delta\mu^2 h^2
+
\delta\lambda h^4
+
\frac{1}{2}\delta m_S^2 S^2
+
\frac{\delta\lambda_S}{4} S^4
+
\frac{\delta\kappa}{2} h^2 S^2 .
\label{eq:Vct_full}
\end{equation}

In principle, all scalar parameters receive one-loop corrections.  
However, in the present analysis we restrict to the background direction $S=0$ and do not impose on-shell renormalization conditions for the singlet sector.  
The singlet parameters $m_{S0}^2$, $\lambda_S$, and $\kappa$ are therefore defined in the $\overline{\rm MS}$ scheme at the renormalization scale $\mu_R$, and no physical pole mass condition is imposed for $S$.

Accordingly, only the Higgs-sector counterterms $\delta\mu^2$ and $\delta\lambda$ are required explicitly in the numerical implementation.  
They are fixed by imposing the renormalization conditions
\begin{align}
\left.\frac{dV_{\rm eff}}{dh}\right|_{h=v,\,S=0,\,T=0} &= 0,
\label{eq:ren_cond1} \\
\left.\frac{d^2V_{\rm eff}}{dh^2}\right|_{h=v,\,S=0,\,T=0} &= m_h^2 ,
\label{eq:ren_cond2}
\end{align}
ensuring that the electroweak vacuum expectation value $v=246$ GeV and the Higgs mass $m_h=125$ GeV remain unchanged at one loop.

Since the singlet does not acquire a vacuum expectation value and no singlet on-shell condition is imposed, the counterterms $\delta m_S^2$, $\delta\lambda_S$, and $\delta\kappa$ do not enter explicitly in the background-field calculation along $S=0$.

With these conditions imposed, the zero-temperature effective potential reproduces the physical Higgs mass and vacuum structure, while the singlet parameters retain their perturbative $\overline{\rm MS}$ definitions at the scale $\mu_R=v$.

\bibliography{ref}

\end{document}